# Encryption and Real Time Decryption for protecting Machine Learning models in Android Applications

*Aryan Verma*

*aryanverma19oct@gmail.com*

*National Institute of Technology, Hamirpur*

## ABSTRACT

*With the Increasing use of Machine Learning in Android applications, more research and efforts are being put into developing better-performing machine learning algorithms with a vast amount of data. Along with machine learning for mobile phones, the threat of extraction of trained machine learning models from application packages (APK) through reverse engineering exists. Currently, there are ways to protect models in mobile applications such as name obfuscation, cloud deployment, last layer isolation. Still, they offer less security, and their implementation requires more effort. This paper gives an algorithm to protect trained machine learning models inside android applications with high security and low efforts to implement it. The algorithm ensures security by encrypting the model and real-time decrypting it with 256-bit Advanced Encryption Standard (AES) inside the running application. It works efficiently with big model files without interrupting the User interface (UI) Thread. As compared to other methods, it is fast, more secure, and involves fewer efforts. This algorithm provides the developers and researchers a way to secure their actions and making the results available to all without any concern.*

**Keywords :-** *Android, Mobile Applications, Security, Data Encryption, Model Deployment*

## 1. Introduction

With the invention of machine learning (ML) techniques for edge computing, mobile phone applications have seen a significant bounce in their usage. Most mobile applications are entirely dependent on the ML models either deployed locally or hosted on the cloud. There are numerous ways in which the trained models are reduced in size and made ready to be used with mobile phones. Organizations and research scientists are putting a lot of effort into developing better algorithms for machine learning and training the models with a vast amount of data. Still, after doing fundamental research and training the model, when the option comes for deploying it to the application, its security also comes into consideration. This work addresses the most vulnerable security threat: extraction of the trained model from the android application. When machine learning models are bundled with the installation file, Android Application Package (APK), or downloaded from the cloud after installing android applications, these model files can be extracted directly after reverse-engineering the APK and used for other third party tasks. This renders the trained model prone to being accessed by someone outside the organization and used without permission.

This work gives an efficient and fast algorithm for securing the trained model files against unauthorized use or decompilation from the APK. The algorithm works in two phases, an encryption phase, and a decryption phase. In the encryption phase, the algorithm encrypts the trained model file with 256-bit AES encryption to a data file that can be bundled with the APK [1][2][3]. Even if the data file is extracted by someone, it will be of no use as it would be encrypted and can't be decompiled or put for generating inference. The second phase is the decryption phase. Wherever there is a need to generate the inference, real-time decryption proposed in work is used, which decrypts the model without storing it in device memory very swiftly. The decryption phase does not interrupt the User interface and works very smoothly in the background. Even big machine learning model files were proven to be decrypted in very little time, which can be neglected while using the application. This prevents the trained model from being used even after the model file in encrypted form is accessed by someone. The proposed algorithm will help the organizations rely less on servers and the cloud for storing their models and investing time to develop APIs (Application Programming Index) in order to access model outputs. The time required to send and receive data also vanishes by deploying the model on the local device, hence the android phone itself.

## 2. Methods to Secure ML Models in Applications

### 2.1 Name Obfuscation

This is the standard practice for most of the classes and resources of the application. Most of the developers use tools to obfuscate the code files in the application [4]. This process converts the file, variable names into another name, making the code difficult to understand. Name Obfuscation of model is to store it with a different filename. This practice ensures very light safety and can easily be breached for the model files in the APK. Hence, someone else can still use the model for any unauthorized task.

### 2.2 Deploying to Cloud

In this method, the model is not stored on the APK, but deployed to the web server. This is a well known practice of securing the machine learning models in mobile applications. Whenever there is a need to generate the inferences from it, an Application programming Interface (API) Call is made from the device to the web server which generates the inference and returns the response of the model to the device. This intakes heavy resources for deploying the model online and handling multiple number of calls. Also this methods renders delay in results as the data transmission timings and concurrency limits of the server creates a time lag. Time, money and expertise of deployment is involved in this method. Many companies and projects use this type of process to secure their models.

### 2.3 Last layer Isolation

This method is used and implemented by very few as it needs a lot of extra effort to be implemented, and it creates a time lag while the model gets ready to be used. The last layer isolation method works by removing the last layer weights from the machine learning model that is to be bundled with APK. In this method, even if the model file is extracted through decompiling the application, it will be of no use as the last layer weights would be absent from the model. Whenever the inference is needed, the last layer weights are fetched either from the server or the application itself. These final layer weights get attached to the model and prepare the entire model file. For getting the inference, we use this model file.

This method takes much extra computation time, and the complexity involved is too high. The generating of inference may take more time due to attaching the last layer or its fetching from the server. Also, if the device stores the entire model, it can be extracted from rooted Android devices. This renders the method more complex and medium-level security to the model file.

### 2.3 Downloading Models on Demand

The trained machine learning model can be downloaded from the back-end during the run time and stored in the app-specific directories. This method doesn't involve storing models in APKs while distributing them for installation, but it downloads the model from the server when the application is put to use by the user. This process fails when the devices are rooted. The model can still be accessed from the folders and can be extracted. Hence, it offers inadequate security to the model file.

### 2.4 Encrytion/Decryption of Model

This method of securing the model file converts the file to an encrypted form which is then decrypted on the client device and ensures maximum safety for the model file in contrast to the other methods [5]. It is applied with less complexity as compared to other methods.

## 3. Overview of Approach

My approach is divided into two parts, first, where the encryption of the model takes place and second which decrypts the model in real-time.

### 3.1 Encryption of Model

This is the first phase of the algorithm. In this, the model is converted to an encrypted data file, which the android application stores and uses in the second phase. The machine learning model file is brought to the android OS or any other system with Java Virtual Machine (JVM). Then the model file is read into the system using Java with the help of Input File Streams. The read file is then converted into Byte Buffers which is due to its fast processing speed and efficient memory usage.

A string variable key is used to initialize the SecretKey, which is a type safe parameter for all further operations to be held in encrypting the model. This SecretKey is used to initialize the encryption algorithm. Here in this work, the AES algorithm is used with a 256-bit key. The Cipher class is used to initialize the AES algorithm.

This initialized class is now put to use by passing the previously read byte array into the instance. This instance works heavily on the byte array form of the model file and encrypts it. The encryption function outputs a byte array which is then stored in the form of a generic data file with a .dat extension. We can use this file to bundle with the APK; even if someone extracts it, it will be of no use until one has the key for its decryption.

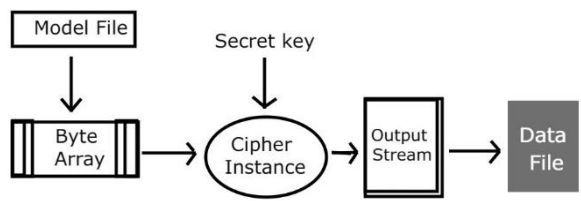

**Figure-1 :-** Overview of Encryption phase of the algorithm. The model file is converted to an encrypted data file using the cipher instance and the key which is stored in the APK.

## 3.2 Decryption and Use

This is the second phase of the algorithm, where the encrypted data file is decrypted and put to use. The encrypted file bundled with the APK is now used for generating the inference in the Application. For this, the file is read within the Application from the assets and converted again to a byte array for further processing. A call is made to fetch the key from the server via the API, which returns a JSON object, and this returned object is parsed to get the key field out of it.

The key field is of the String type and is used to initialize a SecretKeySpec class instance which further is used to initialize the Cipher class in decrypt mode. The encrypted model byte array is then passed to this function which in return gives a byte array. This returned byte array is the original form of the trained model and can be used to generate inferences through the interpreters. This has not to be stored on the device because of the insecurity of decompiling and extraction of file.

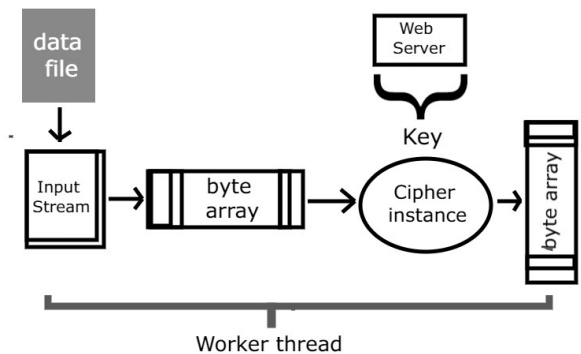

**Figure-2 :-** Overview of the decryption phase of algorithm. The encrypted model in form of data file is again converted to a byte array which is a decrypted version of the machine learning model. This byte can be put to use by passing it to the interpreter and generating the inference.

## 4. Methods

### 4.1 Creating and Storing Key

The key is required for both the encryption and decryption of the model in the Android application. The key is a string data type variable and must be 16 characters long, as the encryption is being performed with 256 bits. The String data type uses 2 bytes for each character, which sums up to 32 bytes for a 16 character long key, which further equals 256 bits. Storing the key inside the APK in the form of BuildConfig Field or in the Native C++ class does not ensure proper protection and the APK can be decompiled for the key. In this work the key is stored on a server which can be returned via an API call with JWT Protection. This ensures that the key is not stored in the APK and is fetched and used dynamically.

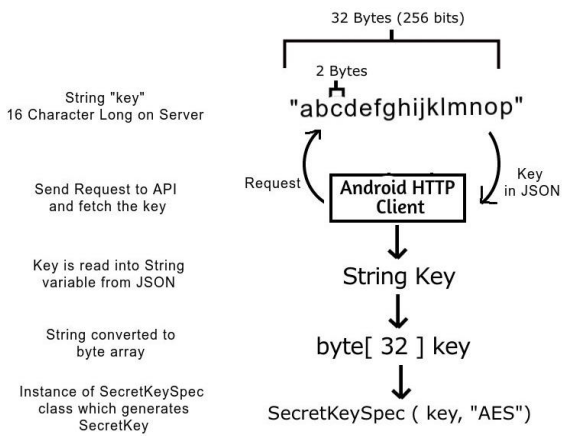

**Figure-3 :-** A 16 character key is stored on the server which is fetched by making a secure call to the server using Java Web Token (JWT) through HTTP Client. The data in form of Java Script Object Notation (JSON) is parsed to a string variable which is then converted to a byte array. This byte array is used in the constructor of SecretKeySpec Class to create its instance.

### 4.2 Generating SecretKeySpec from key

Now, the key, which is in string form, is to be converted into a SecretKey Variable which is to be used for initializing the Cipher. First, the string key variable is converted into a byte array containing the data from the variable. This byte array is then converted to a SecretKey. The SecretKey is an interface from the Java cryptography library providing type safety for parameters that are used for encrypting and decrypting the data. For conversion, SecretKeySpec class is used, which is a helper class, and constructs the SecretKey from the provided byte array without going through a provider-dependent fashion into the android phones. This instance of the SecretKeySpec Class is used to generate the instance of the Cipher class, which is used to encrypt/decrypt the data.

### 4.3 Creating and Initializing the Cipher

Java Cipher is a class from Java Cryptography API, which represents an encryption algorithm, and its instance is used to encrypt/decrypt the data in the application [6][7]. In this work, the cipher instance is used twice, one while creating the encrypted file and the other which is while decrypting the data file back to the model. A parameter tells the cipher class for which type of algorithm is to be used. Here in this work, AES has been used to create the Cipher instance with a 256-bit key. The AES encrypts the model to an extent that it is impossible for the model file to get decrypted back. Even if this AES encrypted file is extracted from the APK, it will be of no use. The machine learning interpreters and the decompilers, both will not be able to read and identify the data.

```
//Fetch and store the key into Key variable
String key ;
String ENC_ALGO = "AES" ;

//Converting the key to byte array
byte[] secretKey = key.getBytes();

//Initializing the Cipher using SecretKeySpec
SecretKeySpec skeySpec = new SecretKeySpec(secretKey, ENC_ALGO);
Cipher cipher = Cipher.getInstance(ENC_ALGO);
cipher.init(Cipher.ENCRYPT_MODE, skeySpec);
```

**Figure-4 :-** Java implementation of the Cipher initialization and secret key generation from the string variable.

### 4.4 Encrypting and Storing Machine Learning Model

The generated instance of the Cipher class is now put to work for encrypting the model file. The model file is read from memory, and data from this file is converted to a byte array. This conversion to a byte array is because the cipher instance accepts the data in the form of a byte array. Now the data as a byte array is passed to the cipher instance for encryption. The output from the instance is received as a byte array, which is an encrypted form of the data we passed to the cipher. This encrypted data has come from the output to be written inside a generic data file. This writing operation is made successful via FileOutputStream. FileOutputStream is a helper class from Java providing a function to write data to a file and save it to a path given to the function.

This FileOutputStream instance writes the data that has been encrypted in the form of a byte array to a generic data file with a .dat extension. This data file is the final encrypted form of the model, which is to be bundled with the APK. After this step, the decryption phase starts, which has to be performed on the device where the APK resides. Even if this data file is accessed through reverse engineering, it will be of no use until one has the key with them.

```
// Read the entire model into a byte array.
InputStream is = getAssets().open("model.tflite");
int size = is.available();
byte[] largeFileBytes = new byte[size];
is.read(largeFileBytes);

//Encrypt the byte array usign cipher instance
byte[] largeFileEncBytes = cipher.doFinal(largeFileBytes);

//Making encrypted file
//This file can be put in APK build
File encrypted = new File("encrypted_model.dat");
FileOutputStream fos = new FileOutputStream(encrypted);
fos.write(largeFileEncBytes);
```

**Figure-4** Java implementation of Encryption and storage of ML model. The model file is read into a byte buffer named largeFileBytes. This buffer is passed to the cipher instance which encrypts it.

### 4.5 Multi-Threading for preventing UI Frame Skipping and Looper Blocking

The decryption of a model for real-time use is a task that consumes up the processor power and may result in the UI thread's stoppage. A multithreading environment is created to prevent the UI from blocking, and decrypting the model in real-time. A new worker thread is started, which is responsible for the whole process of loading and decrypting the model from memory as seen in Figure 5. This prevents skipping the frames and blocking the UI thread.

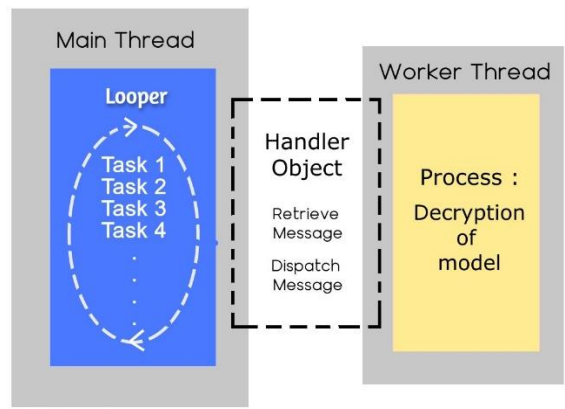

**Figure 5 :-** The heavy process occurring on the worker thread of android application doesn't interrupt the Main UI thread and it works efficiently.

### 4.6 Decrypting Model and use

The encrypted data file to be decrypted and used for generating inference is read from the path where this file is located into a byte array. Byte arrays are used for faster and effective data transmission between the processes; other data types prove to be slower and result in data loss from the model file. Again a cipher is initialized with the same key that we used to encrypt the model file but with the decryption mode. The byte buffer from the encrypted file is passed to the cipher instance. The result is the decrypted data in the form of the byte array. This byte array is the data present in the model file and can be put to a byte buffer and used for initializing the interpreter.

```
//Read the encrypted file in byte array
InputStream i = getAssests().open("encrypted_model.dat");
int size = i.available();
byte[] largeFile = new byte[size];
i.read(largeFile);

//Initialize cipher in decrypt mode
//Decrypt the data
cipher.init(Cipher.DECRYPT_MODE, skeySpec);
byte[] bytes = cipher.doFinal(largeFile);

//Prepare buffer and pass to interpreter
ByteBuffer buffer = ByteBuffer.allocateDirect(model.length)
    .order(ByteOrder.nativeOrder());
buffer.put(bytes);
tensorflow.lite.Interpreter interpreter = new Interpreter(buffer);
```

**Figure-5** Java implementation of Decryption of ML model. The process takes place on a worker thread, which after decrypting the model creates byte array. This array is passed to tensorflow interpreters.

## 5. Results

The algorithm proposed in this work efficiently encrypts well-known architectures in minimal time [8][9][10][11]. The algorithm is tested on various architecture model files with varying trained file sizes on two devices of different CPU processing powers. The files were stored in the form of TensorFlow models converted to flat buffer files using TensorFlow lite library with .tflite extension.

On a mobile device having a Central Processing Unit (CPU) of 8 cores and a clock speed of 2000 MegaHertz (MHz), the U-Squared Net model file with a size of 2560 Kb (Kilobytes) took 227 ms (milliseconds) to get converted to an encrypted byte buffer and 17 ms to get stored in the form of encrypted data (.dat) file. Inception V2, with a size of 11571 Kb, took 720 ms for encrypted byte buffer and 66 ms further to get stored to an encrypted data file. Even big model files such as Inception V3 and EfficientNetB5 took a total of 2079 ms and 1152 ms, respectively, for converting to an encrypted file, giving very fast encryption rates.

**Table-1:** Performance of the encryption algorithm on a mobile device having a CPU of 8 cores and clock speed 2000 MHz.

| Model (tflite) | Size (Megabytes) | Encryption time (milliseconds) | Storage Time (milliseconds) | Total Time (milliseconds) |
|---|---|---|---|---|
| U-Squared Net | 2.5 | 227 | 17 | 244 |
| SSD Mobilenet | 4.2 | 306 | 37 | 343 |
| Inception V2 | 11.3 | 720 | 66 | 786 |
| EfficientNetB5 | 16 | 1054 | 98 | 1152 |
| MnasNet1.0 | 17.5 | 1106 | 101 | 1207 |
| Inception V3 | 23.9 | 1960 | 119 | 2079 |

The algorithm was still giving good results on a mobile device with four cores and a clock speed of 578 MHz. It encrypted the U-squared net model to an encrypted byte buffer in 279 ms and stored it to another file in 33 ms. Inception V2, which took 720 ms in 8 core CPU, took 789 ms on this device to get converted to encrypted byte array and a further 89 ms to get stored to a data file. The storage and encryption of the models were at a pace even for the big model files such as Inception V3 and EfficientNet B5, which took 2304 ms and 1299 ms total for getting converted to an encrypted data file, respectively. This shows that the algorithm is very efficient to encrypt the models on an android device itself.

While using the encrypted model in the application, the decryption timing of the encrypted files proved to be extremely fast in Android OS. On an eight-core CPU device with a clock speed of 2000MHz, the U-squared net and SSD Mobilenet encrypted model files took a total of 203 and 279 ms respectively to get decrypted, which is extremely fast and does not interrupt any routine calls of UI Thread. Inception V2 with an encrypted file size of 11571 Kb took 643 ms to get decrypted. Also, The algorithm decrypted large model files EfficientNet B5 and Inception V3 with an encrypted file size of 16384 Kb and 24473 Kb in 894 ms and 1766 ms, which rendered a very smooth and uninterrupted execution of the Looper thread for UI updates thread for UI updates.

**Table-2:** Performance of the encryption algorithm on a mobile device having a CPU of 4 cores and clock speed 578 MHz.

| Model (tflite) | Size (Megabytes) | Encryption time (milliseconds) | Storage Time (milliseconds) | Total Time (milliseconds) |
|---|---|---|---|---|
| U-Squared Net | 2.5 | 279 | 33 | 312 |
| SSD Mobilenet | 4.2 | 361 | 57 | 418 |
| Inception V2 | 11.3 | 789 | 89 | 878 |
| EfficientNetB5 | 16 | 1176 | 123 | 1299 |
| MnasNet1.0 | 17.5 | 1187 | 131 | 1318 |
| Inception V3 | 23.9 | 2097 | 207 | 2304 |

**Table-3.** Performance of the decryption algorithm on a mobile device having a CPU of 8 cores and clock speed 2000 MHz.

| Model (tflite) | Size (Megabytes) | Decryption time (milliseconds) |
|---|---|---|
| U-Squared Net | 2.5 | 203 |
| SSD Mobilenet | 4.2 | 279 |
| Inception V2 | 11.3 | 643 |
| EfficientNetB5 | 16 | 894 |
| MnasNet1.0 | 17.5 | 933 |
| Inception V3 | 23.9 | 1766 |

Also on quad core device the decryption of the models was very swift and smooth, without interrupting the UI. The U-Squared Net took 266 ms on this device and the SSD mobilenet was decrypted into a byte buffer in 349 ms which proves the algorithm decryption phase to be very swift in small processors also. Big model files were also decrypted in real time on the worker thread of application.

**Table-4:.** Performance of the decryption algorithm on a mobile device having a CPU of 4 cores and clock speed 578 MHz.

| Model (tflite) | Size | Decryption time (ms) |
|---|---|---|
| U-Squared Net | 2.5 Mb | 266 |
| SSD Mobilenet | 4.2 Mb | 349 |
| Inception V2 | 11.3 Mb | 772 |
| EfficientNetB5 | 16 Mb | 1159 |
| MnasNet1.0 | 17.5 Mb | 1151 |
| Inception V3 | 23.9 Mb | 2071 |

## 6. Conclusion

The algorithms proposed have proven to be significantly fast and efficient while encrypting and decrypting the machine learning models. It was tested on multiple devices with varying CPU architectures. Even in mobile phones with less processing powers, the algorithm proved to be very efficient in terms of speed. Even model files having large sizes are also decrypted in real-time using the algorithm. There is no need to store the model on the cloud and put extra effort into API development due to the fear of reverse engineering of APK and extraction of model files from them. Inference timings for any task are also improved as there is no need to send and receive the input data to model through the internet. This method proves to be very much secure from obfuscation and last layer isolation. The cost and efforts to implement the algorithm was also very less as compared to the last layer isolation and placing the model on web servers

## 7. Future Work

There may be a need to put some very large model files into the APK and use the proposed algorithm to secure them. In those cases, it may produce some time lag while decrypting the model. This process can be made faster. The model file can be fragmented into multiple files while encrypting it. The buffer coming from the encryption cipher can be transferred to different small files, and these files can be bundled with APK. These fragments can be decrypted separately on parallel running threads and added to a single byte buffer which becomes the final model and takes care of initializing the interpreter. Processing the fragments parallelly on different worker threads will render very fast decryption of the encrypted model file.

If the key of the model file can be stored on the android device itself, it would be better, as there would be no need for any server or API calls. The technique to store the key on the device itself in terms of the build config field or in the C++ files may be breached. There has to be a method for storing the key in the APK so that it doesn't show up on decompiling.up.

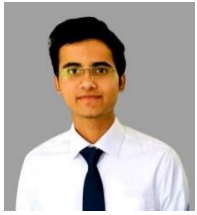

**Aryan Verma**
Student, NIT Hamirpur

He is currently pursuing his Bachelors with major as Computer Science from National institute of Technology, Hamirpur, India. He has been to various roles of software engineering and currently is a Google Summer of Code 2021 student with Department of Biomedical Informatics (BMI), Emory university School of Medicine. His research interests are Computer Vision, Mobile Computing and Machine Learning.